\documentclass[aps,prl,superscriptaddress,showpacs,twocolumn,floatfix]{revtex4}

\usepackage{graphicx}

\def\deg{^\circ}
\def\gtorder{\mathrel{\raise.3ex\hbox{$>$}\mkern-14mu
 \lower0.6ex\hbox{$\sim$}}}
\def\ltorder{\mathrel{\raise.3ex\hbox{$<$}\mkern-14mu
 \lower0.6ex\hbox{$\sim$}}}
\def\mugegm{\mu_p G_E / G_M}
\def\gegm{G_E / G_M}
\def\ge{G_E}
\def\gm{G_M}
\def\gd{G_D}

\begin{document}

\title{Precision Rosenbluth measurement of the proton elastic form factors}

\author{I.~A.~Qattan}
\affiliation{Northwestern University, Evanston, Illinois, 60208}
\author{J.~Arrington}
\affiliation{Argonne National Laboratory, Argonne, Illinois, 60439}
\author{R.~E.~Segel}
\affiliation{Northwestern University, Evanston, Illinois, 60208}
\author{X.~Zheng}
\affiliation{Argonne National Laboratory, Argonne, Illinois, 60439}

\author{K.~Aniol}
\affiliation{California State University, Los Angeles, Los Angeles, California, 90032}
\author{O.~K.~Baker}
\affiliation{Hampton University, Hampton, Virginia, 23668}
\author{R.~Beams}
\affiliation{Argonne National Laboratory, Argonne, Illinois, 60439}
\author{E.~J.~Brash}
\affiliation{University of Regina, Regina, Saskatchewan, Canada, S4S 0A2}
\author{J.~Calarco}
\affiliation{University of New Hampshire, Durham, New Hampshire, 03824}
\author{A.~Camsonne}
\affiliation{Universit\'{e} Blaise Pascal Clermont-Ferrand et CNRS/IN2P3 LPC 63, 177 Aubi\`{e}re Cedex, France} 
\author{J.-P.~Chen}
\affiliation{Jefferson Laboratory, Newport News, Virginia, 23606}
\author{M.~E.~Christy}
\affiliation{Hampton University, Hampton, Virginia, 23668}
\author{D.~Dutta}
\affiliation{Massachusetts Institute of Technology, Cambridge, Massachusetts, 02139}
\author{R.~Ent}
\affiliation{Jefferson Laboratory, Newport News, Virginia, 23606}
\author{S.~Frullani}
\affiliation{Istituto Nazionale di Fisica Nucleare, Sezione Sanit\`{a}, 00161 Roma, Italy}
\author{D.~Gaskell}
\affiliation{University of Colorado, Boulder, Colorado, 80309}
\author{O.~Gayou}
\affiliation{College of William and Mary, Williamsburg, Virginia, 23187}
\author{R.~Gilman}
\affiliation{Rutgers, The State University of New Jersey, Piscataway, New Jersey, 08855}
\affiliation{Jefferson Laboratory, Newport News, Virginia, 23606}
\author{C.~Glashausser}
\affiliation{Rutgers, The State University of New Jersey, Piscataway, New Jersey, 08855}
\author{K.~Hafidi}
\affiliation{Argonne National Laboratory, Argonne, Illinois, 60439}
\author{J.-O.~Hansen}
\affiliation{Jefferson Laboratory, Newport News, Virginia, 23606}
\author{D.~W.~Higinbotham}
\affiliation{Jefferson Laboratory, Newport News, Virginia, 23606}
\author{W.~Hinton}
\affiliation{Old Dominion University, Norfolk, Virginia, 23529}
\author{R.~J.~Holt}
\affiliation{Argonne National Laboratory, Argonne, Illinois, 60439}
\author{G.~M.~Huber}
\affiliation{University of Regina, Regina, Saskatchewan, Canada, S4S 0A2}
\author{H.~Ibrahim}
\affiliation{Old Dominion University, Norfolk, Virginia, 23529}
\author{L.~Jisonna}
\affiliation{Northwestern University, Evanston, Illinois, 60208}
\author{M.~K.~Jones}
\affiliation{Jefferson Laboratory, Newport News, Virginia, 23606}
\author{C.~E.~Keppel}
\affiliation{Hampton University, Hampton, Virginia, 23668}
\author{E.~Kinney}
\affiliation{University of Colorado, Boulder, Colorado, 80309}
\author{G.~J.~Kumbartzki}
\affiliation{Rutgers, The State University of New Jersey, Piscataway, New Jersey, 08855}
\author{A.~Lung}
\affiliation{Jefferson Laboratory, Newport News, Virginia, 23606}
\author{D.~J.~Margaziotis}
\affiliation{California State University, Los Angeles, Los Angeles, California, 90032}
\author{K.~McCormick}
\affiliation{Rutgers, The State University of New Jersey, Piscataway, New Jersey, 08855}
\author{D.~Meekins}
\affiliation{Jefferson Laboratory, Newport News, Virginia, 23606}
\author{R.~Michaels}
\affiliation{Jefferson Laboratory, Newport News, Virginia, 23606}
\author{P.~Monaghan}
\affiliation{Massachusetts Institute of Technology, Cambridge, Massachusetts, 02139}
\author{P.~Moussiegt}
\affiliation{Laboratorie de Physique Subatomique et de Cosmologie, F-38026 Grenoble, France}
\author{L.~Pentchev}
\affiliation{College of William and Mary, Williamsburg, Virginia, 23187}
\author{C.~Perdrisat}
\affiliation{College of William and Mary, Williamsburg, Virginia, 23187}
\author{V.~Punjabi}
\affiliation{Norfolk State University, Norfolk, Virginia, 23529}
\author{R.~Ransome}
\affiliation{Rutgers, The State University of New Jersey, Piscataway, New Jersey, 08855}
\author{J.~Reinhold}
\affiliation{Florida International University, Miami, Florida, 33199}
\author{B.~Reitz}
\affiliation{Jefferson Laboratory, Newport News, Virginia, 23606}
\author{A.~Saha}
\affiliation{Jefferson Laboratory, Newport News, Virginia, 23606}
\author{A.~Sarty}
\affiliation{Saint Mary's University, Halifax, Nova Scotia, Canada B3H 3C3}
\author{E.~C.~Schulte}
\affiliation{Argonne National Laboratory, Argonne, Illinois, 60439}
\author{K.~Slifer}
\affiliation{Temple University, Philadelphia, Pennsylvania, 19122}
\author{P.~Solvignon}
\affiliation{Temple University, Philadelphia, Pennsylvania, 19122}
\author{V.~Sulkosky}
\affiliation{College of William and Mary, Williamsburg, Virginia, 23187}
\author{K.~Wijesooriya}
\affiliation{Argonne National Laboratory, Argonne, Illinois, 60439}
\author{B.~Zeidman}
\affiliation{Argonne National Laboratory, Argonne, Illinois, 60439}

\date{\today}

\begin{abstract}

We report the results of a new Rosenbluth measurement of the proton
form factors at $Q^2$ values of 2.64, 3.20 and 4.10 GeV$^2$.  Cross
sections were determined by detecting the recoiling proton in contrast to
previous measurements in which the scattered electron was detected.  At each
$Q^2$, relative cross sections were determined to better than 1\%.  The
measurement focussed on the extraction of $\gegm$ which was determined to
4--8\% and found to approximate form factor scaling, \textit{i.e.} $\mu_p \ge
\approx \gm$. These results are consistent with and much more precise than
previous Rosenbluth extractions.  However, they are inconsistent with recent
polarization transfer measurements of comparable precision, implying a
systematic difference between the two techniques.

\end{abstract}

\maketitle

%%%%%%%%%%%%%%%%%%%%%%%%%%%%%%%%%%%%%%%%%%%%%%%%%%%%%%%%%%%%%%%%%%%%%%%%%%%%%

Reproducing the structure of the proton is one of the defining problems of QCD.
The electromagnetic structure can be expressed in terms of the electric and
magnetic form factors, $\ge$ and $\gm$, which  depend only on the 4-momentum
transfer squared, $Q^2$. They have traditionally been determined utilizing
the Rosenbluth formula~\cite{rosenbluth50} for elastic $e$--$p$ scattering:
\begin{equation}
\frac{d\sigma}{d\Omega_e} = \frac{\sigma_{Mott}}{\varepsilon(1+\tau)}
\bigl[ \tau \gm^2(Q^2) + \varepsilon \ge^2(Q^2) \bigr],
\label{eqn:sig}
\end{equation}
where $\tau=Q^2/4M_p^2$, $\varepsilon$ is the longitudinal polarization
of the exchanged virtual photon, $\varepsilon^{-1} =
1+2(1+\tau)\tan^2{(\theta_e/2)}$, $M_p$ is the proton mass, and $\theta_e$ is
the electron scattering angle.  The form factors are related to the spatial
distributions of the charge ($\ge$) and magnetization ($\gm$) in the proton,
and in the non-relativistic limit are simply the Fourier transformations of
these distributions.

A Rosenbluth separation is performed by varying the incident electron energy
and electron scattering angle to keep $Q^2$ constant while varying
$\varepsilon$. Rosenbluth separations of $\ge$ and $\gm$ have been
reported from 1960 to the present day (see Refs.~\cite{walker94, christy04}
and references therein). Fits to these data yield $\mugegm \approx
1$~\cite{walker94, arrington03a}, implying similar charge and magnetization
distributions. At large $Q^2$ values, $\gm$ dominates the cross section at all
$\varepsilon$ values (contributing more than 90\% for $Q^2 > 4$~GeV$^2$), and
thus while a Rosenbluth separation can yield a precise extraction of $\gm$,
the uncertainty in $\ge$ increases with increasing $Q^2$.

The high-$Q^2$ behavior of the electric form factor can be more precisely
determined in polarization transfer experiments, where longitudinally
polarized electrons are scattered from unpolarized protons and both transverse
and longitudinal polarization are transferred to the struck proton. During the
last few years, polarization transfer experiments have been performed at
Jefferson Lab which measured $\gegm$ up to $Q^2$ = 5.6~GeV$^2$~\cite{jones00,
gayou02}. These measurements show the ratio decreasing with increasing $Q^2$
in stark contrast to the approximate scaling observed in Rosenbluth
measurements.

At high $Q^2$, the quoted uncertainties on the polarization transfer results
are much smaller than those for the Rosenbluth extractions. This fact,
combined with the scatter in the results from different Rosenbluth
measurements, led to speculation that the Rosenbluth determinations of $\ge$
were unreliable. While the scatter appears to be the result of incomplete
treatment of normalization uncertainties when combining data from different
experiments~\cite{arrington03a}, the Rosenbluth technique is still very
sensitive to small corrections to the cross section at large $Q^2$ values. In
this letter, we report the results of a new experiment that utilizes an
improved Rosenbluth technique to determine the proton form factor ratio
$\gegm$ with uncertainties a factor of two to three smaller than all previous
Rosenbluth measurements and comparable to the precision of the polarization
transfer measurements.

Experiment E01-001 was performed in Hall A at Jefferson Lab. A 70~$\mu$A
electron beam with energies from 1.9 to 4.7 GeV impinged on a 4-cm liquid
hydrogen (LH2) target.  Protons from elastic $e$--$p$ scattering were detected
in the High Resolution Spectrometer. An aerogel Cerenkov detector was used to
eliminate charged pions from the data, and data from a ``dummy'' target were
used to measure the contribution from the aluminum walls of the target.
Table~\ref{tab:kine} lists the kinematics of the experiment. Detailed
descriptions of the spectrometer and beamline instrumentation can be found in
Ref.~\cite{alcorn04}.

\begin{table}[htb]
\begin{center}
\caption{Kinematics of the experiment.
\label{tab:kine}}
\begin{tabular}{|c|c|c|c|c|c|c|}
\hline
$E_{beam}$ & \multicolumn{2}{|c|}{$Q^2$=2.64~GeV$^2$} & \multicolumn{2}{c|}{$Q^2$=3.20~GeV$^2$} & \multicolumn{2}{c|}{$Q^2$=4.10~GeV$^2$} \\
 (GeV)	& $\varepsilon$ & $\theta_p$ ($\deg$) & $\varepsilon$ & $\theta_p$ ($\deg$) & $\varepsilon$ & $\theta_p$ ($\deg$) \\
\hline
1.912 	& 0.117 & 12.631	& --	& --		& -- & --	\\
2.262 	& 0.356 & 22.166	& 0.131 & 12.525	& -- & --	\\
2.842 	& 0.597 & 29.462	& 0.443 & 23.395	& 0.160 & 12.682 \\
3.772 	& 0.782 & 35.174	& 0.696 & 30.500	& 0.528 & 23.666 \\
4.702 	&~0.865~& 38.261	&~0.813~& 34.139	&~0.709~& 28.380 \\
\hline
\end{tabular}
\end{center}
\end{table}

%The table is fully self-consistent, good to three decimal places (see test_kine.pcm)

Because the fractional contribution of $\ge$ to the cross section is small at
large $Q^2$, its extraction is highly sensitive to corrections that modify
the $\varepsilon$ dependence. All previous Rosenbluth separations involved
detection of the electron. The rapid variation of the electron cross section
with scattering angle means that any rate-dependent corrections to the cross
section will have a strong $\varepsilon$ dependence. Varying $\varepsilon$
also leads to large changes in the momentum of the scattered electron, and so
momentum-dependent corrections will affect the extracted value of $\ge$.
Finally, radiative corrections to the cross section also yield significant
$\varepsilon$-dependent corrections.

Detecting the struck proton greatly reduces all of these sources of
uncertainty. The proton differential cross section ($d\sigma/d\Omega_p$)
changes by less than a factor of two over the $\varepsilon$ range of the
measurement while the electron cross section ($d\sigma/d\Omega_e$) varies by
almost two orders of magnitude. In addition, the minimum cross section is
twenty times larger for the proton than for the electron. The proton momentum
is constant at fixed $Q^2$, while the corresponding electron momentum varies
by a factor of ten. Additionally, the corrections due to beam energy offsets,
scattering angle offsets, and radiative corrections (in particular the
electron bremsstrahlung corrections) all have a smaller $\varepsilon$
dependence when the proton is detected.

\begin{figure}
\includegraphics[height=4.4cm,width=8.0cm,angle=0]{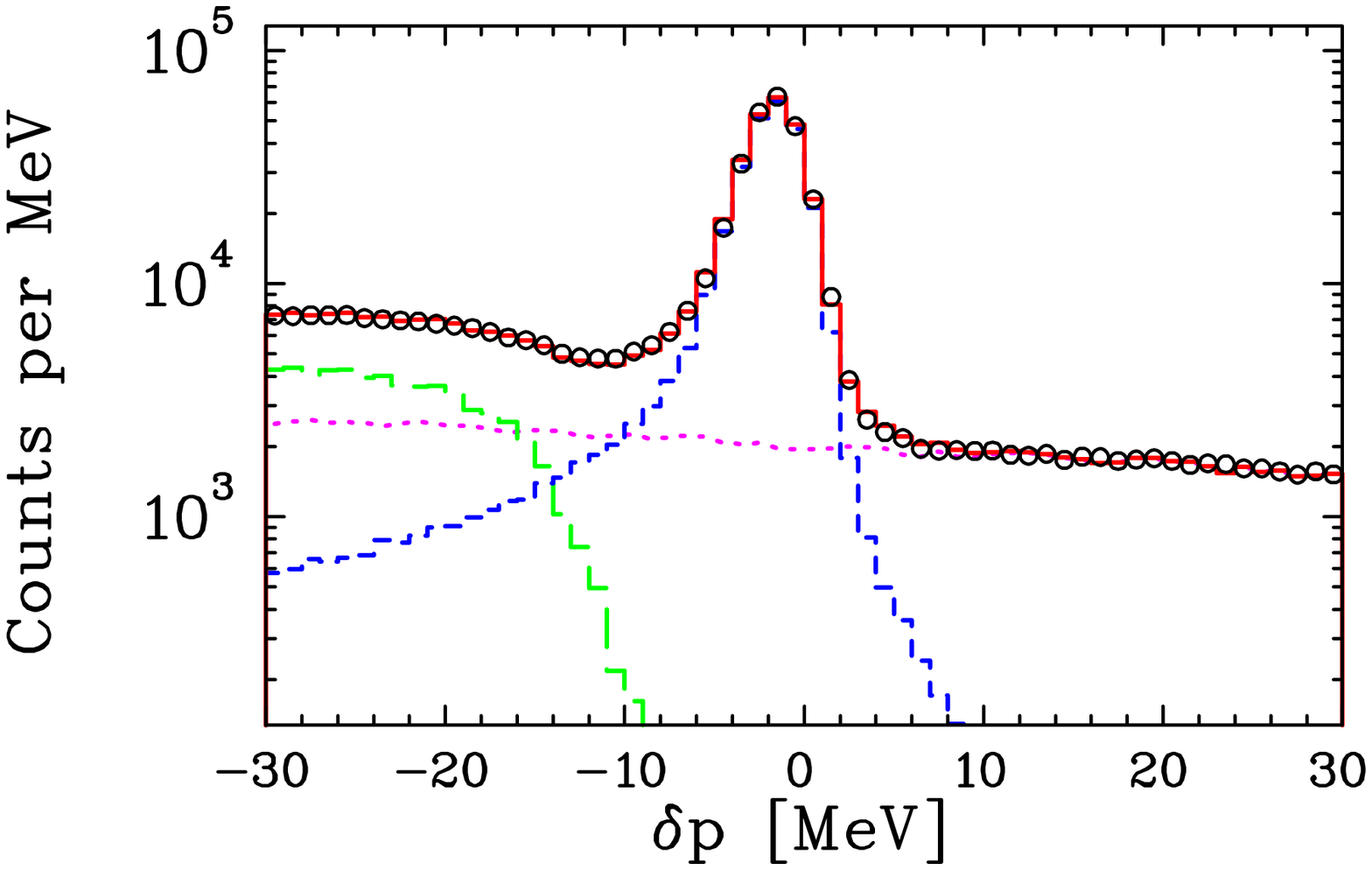}
\includegraphics[height=4.4cm,width=8.0cm,angle=0]{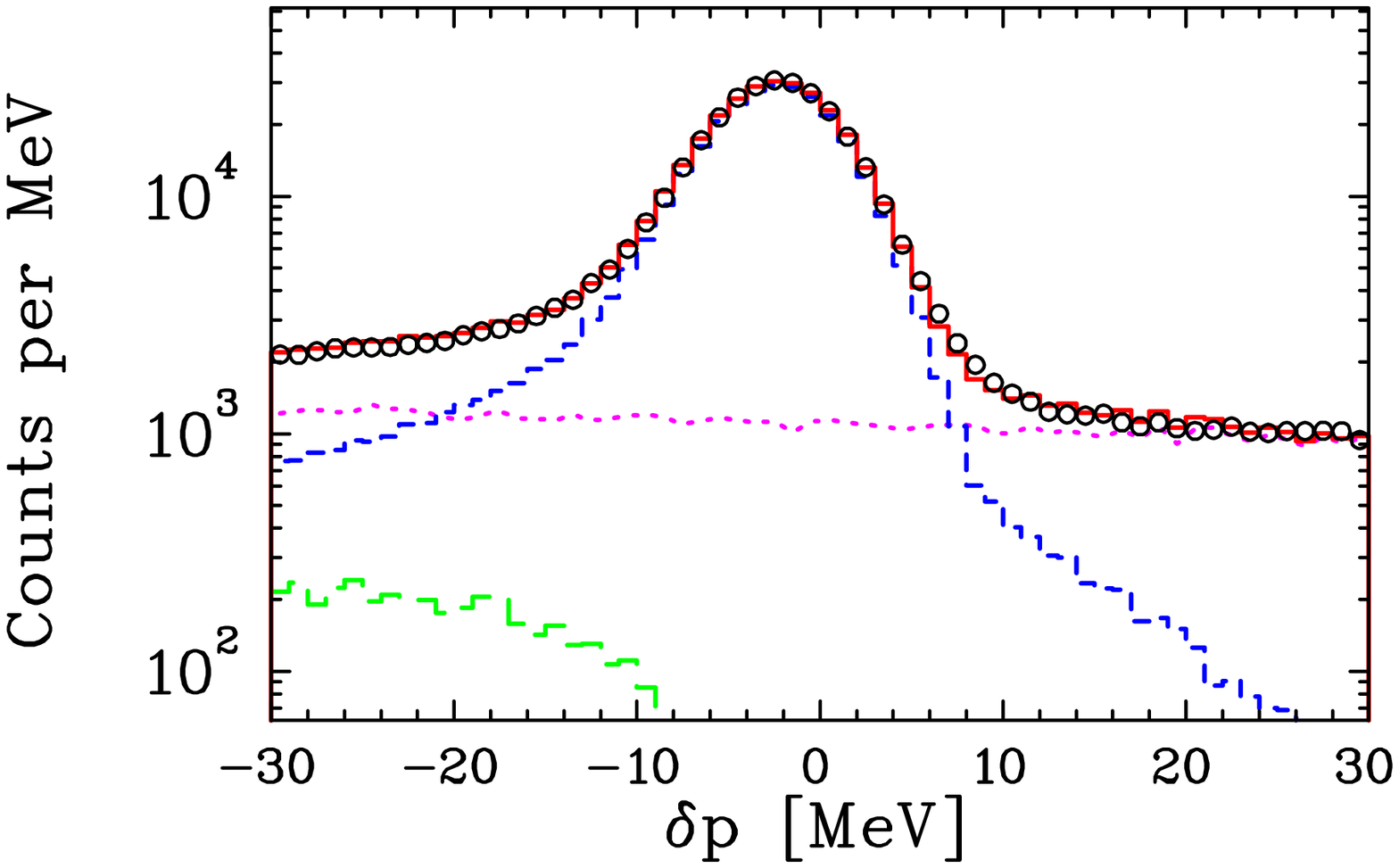}
\caption{(Color online) The measured $\delta p$ spectrum for the low (top) and
high (bottom) $\varepsilon$ points at $Q^2=3.2$~GeV$^2$ (circles).  The dotted
(magenta) line is the background from the target walls, the long-dash line
(green) is the simulated background from $\gamma  p \rightarrow \pi^0 p$ and
$\gamma p \rightarrow \gamma p$ reactions, the short-dash (blue) line is the
simulated elastic spectrum, and the solid (red) line that goes through the
data is the sum of the target wall, elastic, and background contributions
after each contribution is normalized to the data (see text).
\label{fig:deltap}}
\end{figure}

Figure~\ref{fig:deltap} shows the ``missing momentum'' spectra at two
$\varepsilon$ values for $Q^2 = 3.20$~GeV$^2$. The missing momentum, $\delta
p$, is defined as the difference between the measured proton momentum and the
proton momentum expected for elastic scattering at the measured proton angle.
The $\delta p$ spectrum is dominated by the $e$--$p$ elastic peak at $\delta p
\approx 0$. To precisely separate elastic scattering from other processes, we
need to know the shape of the elastic peak. The peak is simulated using the
Monte Carlo code SIMC which takes into account the acceptance and resolution
of the spectrometer, energy loss and multiple scattering of the proton, and
radiative corrections~\cite{ent01}. The resolution of the simulation has been
modified to reproduce the small non-gaussian tails observed in the data. These
are matched to the coincidence data, taken for two beam energies at
$Q^2$=2.64~GeV$^2$, where the background contributions that dominate the
singles spectrum at large $\mid$$\delta p$$\mid$ values are strongly
suppressed. The resolution of the elastic peak is dominated by the angular
resolution. The greater width at large $\varepsilon$ is due to the increased
sensitivity of proton momentum to scattering angle.

Also shown in Fig.~\ref{fig:deltap} is the decomposition of the background
into two components. The background that extends to high $\delta p$ is due to
quasielastic scattering and other reactions in the target walls.  The other
background is mainly due to $\gamma p \rightarrow \pi^0 p$ events, with a 
small (1--4\%) contribution from $\gamma p \rightarrow \gamma p$.  The
spectrum from these reactions was modeled using a calculated bremsstrahlung
spectrum and an $s^{-7}$ cross section dependence. Because of the finite pion
mass, the proton spectrum from pion photoproduction cuts off approximately 10
MeV below the elastic peak.  For small proton angles, where we have the best
resolution, the background from pion photoproduction can be cleanly separated
from the elastic protons while at large angles the background becomes small.
The correction to the elastic cross section due to contributions from the
target walls is approximately 10\%, while the inelastic processes from
hydrogen contribute less than 2\%.

Because the thicknesses are different for the LH2 and dummy targets, the
bremsstrahlung yields are also slightly different.  We use the dummy data to
determine the shape of the endcap contributions, but normalize the
contribution to the LH2 spectrum at large $\delta p$, where the
contribution from the hydrogen is negligible.  While the shape of the
bremsstrahlung spectrum differs slightly between the dummy and LH2 targets,
the effect is only noticeable near the endpoint.  The small uncertainty
due to the difference in shape is taken into account in the systematic
uncertainties.

After removing the background due to the endcaps, the simulated $\pi^0$
photoproduction spectra were normalized to the low-momentum sides of the
$\delta p$ spectra (taking into account the elastic radiative tail).  Removing
this background yields clean spectra of elastic events which are then compared
to the elastic simulation. The elastic cross section is taken to be the value
used in the simulation, scaled by the ratio of counts in the data to counts in
the simulated spectrum.

\begin{figure}
\includegraphics[height=8.5cm,width=8.0cm,angle=0]{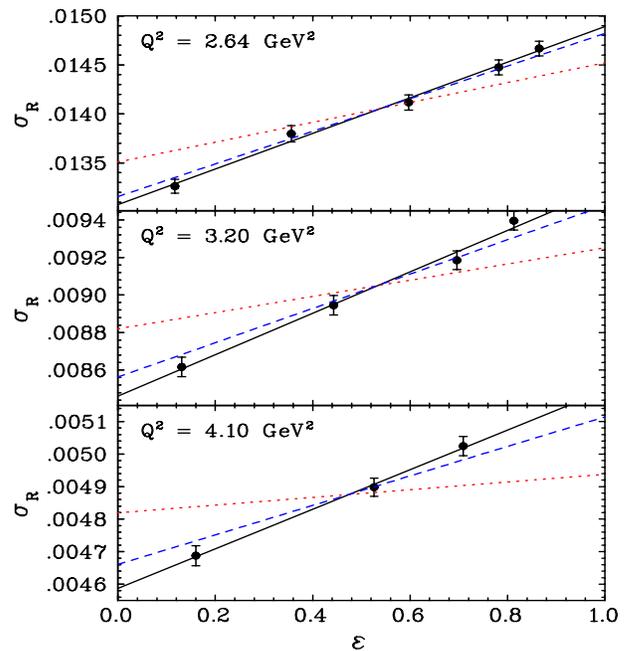}
\caption{(Color online) Reduced cross sections as a function of $\varepsilon$.
The solid line is a linear fit to the reduced cross section,
the dashed line shows the slope predicted from a global analysis of previous
Rosenbluth results~\cite{arrington04a}, and the dotted line shows the slope
predicted by the polarization transfer experiments~\cite{gayou02}.
\label{fig:sigr}}
\end{figure}

The proton yield is corrected for deadtime in the data acquisition system
(10--20\%, measured to better than 0.1\%) as well as several other small
inefficiencies. Corrections for tracking efficiency, trigger efficiency, and
particle identification cuts (using the aerogel Cerenkov detector) were small
($<$2\%) and independent of $\varepsilon$.  Finally, we required a single
clean cluster of hits in each drift chamber plane to avoid events where the
resolution is worsened by noise in the chambers. This significantly reduced
the non-gaussian tails in the reconstructed quantities, but led to an
inefficiency of roughly 7\%, with a small (0.25\%) $\varepsilon$ dependence,
possibly related to the small variation of rate with $\varepsilon$.  We
corrected the yield for the observed inefficiency and applied a 100\%
uncertainty on the $\varepsilon$-dependence of the correction.

The absolute uncertainty on the extracted cross sections is approximately 3\%,
dominated by the uncertainty in the angular acceptance (2\%), the radiative
corrections (1\%), corrections for proton absorption in the target and
detector stack (1\%), the subtraction of endcap and inelastic processes (1\%),
and the uncertainty in the integrated luminosity (1\%). We apply a tight cut
on the solid angle, using only the data in the central 1.6 msr of the total
$\approx$6 msr acceptance. These tight cuts limit the elastic data to the
region of 100\% acceptance, but lead to the relatively large uncertainty in
the size of the software-defined solid angle.  Because the solid angle is
identical for all $\varepsilon$ values at each $Q^2$, this uncertainty affects
the absolute cross section, but not the extraction of $\gegm$.

The largest point-to-point uncertainties, where the error can differ at
different $\varepsilon$ values, are related to the tracking efficiency
(0.2\%), uncertainty in the scattering angle (0.2\%), the subtraction of the
inelastic proton backgrounds (0.2\%), and the radiative corrections (0.2\%).
The total point-to-point systematic uncertainty is 0.45\%, and the typical
statistical uncertainty varies from 0.25\% at $Q^2=2.64$~GeV$^2$ to 0.40\% at
$Q^2=4.1$~GeV$^2$.  The cross sections measured at 2.262 GeV have a slight
additional uncertainty (0.3\%) because these data were taken at lower
beam currents (30-50~$\mu$A), and so are sensitive to any non-linearity in the
beam current measurements and have somewhat different corrections for target
heating.

The reduced cross section, $\sigma_R = \tau \gm^2 + \varepsilon \ge^2$,
is shown as a function of $\varepsilon$ in Fig.~\ref{fig:sigr}. The 
uncertainties shown are the statistical and point-to-point systematic
uncertainties.  Additional corrections, \textit{e.g.} the effect of a fixed
angular offset for all points, would lead to a change in the cross sections
that would vary approximately linearly with $\varepsilon$.  These uncertainties
would not contribute to the scatter of the points or deviations from linearity
in the reduced cross section, but can modify the extracted value of the slope.
These uncertainties are dominated by uncertainties in the $\varepsilon$
dependence of the radiative corrections (0.3\%), background subtraction,
(0.25\%), rate-dependent corrections in the tracking efficiency (0.25\%), and
the effect of a small beam energy or scattering angle offset (0.25\%). These
corrections yield a 0.55\% uncertainty in the slope of the reduced cross
section which is included in the uncertainties of the extracted form factors.

\begin{figure}
\includegraphics[height=5.5cm,width=8.0cm,angle=0]{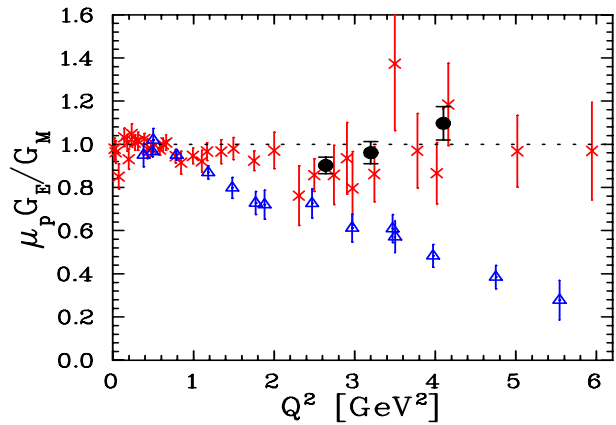}
\caption{(Color online) Extracted values of $\mugegm$ from this work
(circles), a global analysis of previous cross section data (Fig.~2 of
Ref~\cite{arrington04a}) (crosses), and high-$Q^2$ polarization transfer
measurements~\cite{jones00,gayou02} (triangles).
\label{fig:gegm}}
\end{figure}

The form factors are obtained from a linear fit to the reduced cross sections.
The results are given in Table~\ref{tab:gegm} and the ratio $\mugegm$ is
shown in Fig.~\ref{fig:gegm} together with the results of previous Rosenbluth
and polarization transfer measurements.  Note that for consistency with
previous Rosenbluth measurements, the effects of Coulomb
distortion~\cite{arrington04c} have not been included in the results in
Table~\ref{tab:gegm}.  Correcting for Coulomb distortion would lower $\mugegm$
by 0.048, 0.042, and 0.032 and increase $\gm/(\mu_p\gd)$ by 0.009, 0.007, and
0.006 for $Q^2$=2.64, 3.2, and 4.1 GeV$^2$, respectively.

\begin{table}[htb]
\begin{center}
\caption{Form factor values extracted from this measurement relative to the
dipole form, $\gd = 1/(1+Q^2/0.71)^2$.
\label{tab:gegm}}
\begin{tabular}{|l|c|c|c|}\hline
$Q^2$		& 2.64	GeV$^2$		& 3.20	GeV$^2$		& 4.10 GeV$^2$	\\ \hline
$\ge/\gd$	& $0.949\pm0.040$	& $1.007\pm0.052$	& $1.132\pm0.077$ \\
$\gm/(\mu_p\gd)$& $1.053\pm0.015$	& $1.048\pm0.015$	& $1.031\pm0.015$ \\
$\mugegm$	& $0.902\pm0.038$	& $0.961\pm0.051$	& $1.097\pm0.077$ \\
\hline
\end{tabular}
\end{center}
\end{table}

% Preliminary numbers:
%$\mugegm$	& $0.893\pm0.040$	& $0.950\pm0.054$	& $1.081\pm0.084$ \\
%$\ge/\gd$	& $0.938\pm0.042$	& $0.994\pm0.056$	& $1.113\pm0.084$ \\
%$\gm/(\mu_p\gd)$& $1.050\pm0.016$	& $1.045\pm0.016$	& $1.029\pm0.016$ \\

The results presented here are in good agreement with form factors extracted
from previous cross section data, but have much smaller statistical and
systematic uncertainties.  This fact, combined with the consistency of the
various Rosenbluth measurements~\cite{arrington03a}, rules out most
explanations of the discrepancy in terms of possible experimental error in the
cross section measurements.  The accuracy of the present work leaves little
room for doubting that the $\gegm$ ratios reported from the Jefferson Lab
polarization transfer experiments are inconsistent with the form factors
obtained using the Rosenbluth technique, and makes it clear that the problem
is not simply experimental error in previous Rosenbluth measurements. The
source of this discrepancy must be identified before the new insight into the
proton structure provided by the recent polarization transfer data can be fully
accepted.  

One possible source for the difference between the two techniques is the
effect of higher-order processes, or radiative corrections to the Born
(one-photon exchange) cross section. The form factors are extracted from the
cross section (or polarization transfer) measurements assuming the Born
approximation, so the effects of additional processes must be removed from
the measured cross sections. We correct the data for higher-order
electromagnetic interactions such as bremsstrahlung, vertex corrections, and
loop diagrams~\cite{ent01}.  If we had detected the electron, the
bremsstrahlung correction to the $\varepsilon$ dependence of $\sigma_R$ would
have exceeded the $\varepsilon$ dependence coming from $\ge$. In this
experiment, the $\varepsilon$-dependent correction is much smaller and of the
opposite sign. As the other radiative correction terms have almost no
$\varepsilon$ dependence, the consistency between the new data and previous
Rosenbluth results provides a significant verification of the validity of the
standard radiative correction procedures.

It has been suggested that higher-order processes such as two-photon exchange,
not fully treated in standard radiative correction procedures, could explain
the discrepancy~\cite{guichon03, blunden03}.  Such a correction
would have to increase the cross section at large-$\varepsilon$ by
roughly 6\% relative to the low-$\varepsilon$ values.~\cite{arrington04a}.
This would mean that the Rosenbluth form factors, the form factors extracted
using only the cross section data, would have large errors due
to these missing corrections. While two-photon exchange will also affect
the polarization transfer data, the corrections to the polarization transfer
form factors are expected to be smaller, though not necessarily
negligible~\cite{guichon03, arrington04d}.

Additional experimental and theoretical effort is necessary to determine 
if the discrepancy in the form factor measurements can be explained entirely
by higher-order radiative corrections. The effect of multiple soft photon
exchange (Coulomb distortion) has been examined~\cite{arrington04c} and yields
a change in the slope of 1--2\%, corresponding to a 3--5\% reduction in the
extracted value of $\mugegm$, well below the level necessary to explain the
discrepancy.  Recent calculations~\cite{blunden03, chen04} show significant
corrections to the cross sections due to two-photon exchange, though they
appear to explain only half of the observed discrepancy. Additional effort is
going into calculations of two-photon exchange for the cross section and
polarization transfer measurements. It remains to be seen whether the effects
of two-photon exchange are the reason for what must otherwise be considered a
severe discrepancy.

If missing radiative correction terms are responsible for the discrepancy,
then the Rosenbluth form factors, and to a lesser extent the polarization
transfer form factors, will not correspond to the true form factors of the
proton.  The discrepancy must be resolved before precise comparisons
can be made between models of proton structure and the measured form
factors.  While the Rosenbluth data may not provide the true form factors of
the proton, they still provide a useful parameterization of the
electron-proton cross section, with the missing higher-order corrections
absorbed into the extracted form factors. Thus, the Rosenbluth form factors do
reproduce the observed cross sections and provide the best parameterization
when elastic scattering is used to compare the normalization of different
experiments, or when the elastic cross section is used as input to the
analysis of experiments such as quasielastic $A(e,e'p)$
scattering~\cite{dutta03, arrington04a}.

In conclusion, we have performed an improved Rosenbluth extraction of the
proton electromagnetic form factors using detection of the struck proton
rather than the scattered electron to decrease dramatically the uncertainty in
the extraction. The results are as precise as recent polarization transfer
measurements, but are in agreement with previous Rosenbluth separations and
inconsistent with high-$Q^2$ polarization transfer data. The precision
of these new results rules out experimental error in the Rosenbluth
results as the source of the discrepancy between the two techniques, and
provides a stringent test of the radiative corrections that are currently
used in elastic $e$--$p$ scattering. There are indications that this
difference might come from two-photon exchange corrections, but we must better
understand the discrepancy before precise knowledge of the proton form factors
can be claimed.

\begin{acknowledgments}

We gratefully acknowledge the staff of Accelerator Division, the Hall A
technical staff, and the members of the survey and cryotarget groups at
Jefferson Lab for their efforts in making this experiment possible. This work
was supported in part by DOE contract W-31-109-ENG-38, NSF grants 0099540 and
PHY-00-98642, NSERC (Canada), and DOE contract DE-AC05-84ER40150, under
which the Southeastern Universities Research Association operates the
Thomas Jefferson National Accelerator Facility.

\end{acknowledgments}

\bibliography{prl_e01001}

\end{document}